# Comprehensive Investigation and Evaluation of an Indoor 3D System Performance Based on Visible Light Communication

Vailet Hikmat Faraj Al Khattat
Wireless and Photonics Networks
Research Center (WiPNET),
Department of Computer and
Communication Systems Engineering,
Faculty of Engineering
Universiti Putra Malaysia, 43400
UPM Serdang, Selangor, Malaysia
eng.vailet@gmail.com

Siti Barirah Ahmad Anas
Wireless and Photonics Networks
Research Center (WiPNET),
Department of Computer and
Communication Systems Engineering,
Faculty of Engineering
Universiti Putra Malaysia, 43400
UPM Serdang, Selangor, Malaysia
barirah@upm.edu.my

Abdu Saif
Department of Communication and
Computer Engineering, Faculty of
Engineering and IT, Taiz University,
Taiz P.O. Box 6803, Yemen
abdu.saif@taiz.edu.ye

*Abstract*— The obvious and accelerator trend towards efficient green technology in a modern technological revolution time makes visible light communication (VLC) a solution key to meeting this growth. Besides the illumination function, VLC uses for data transmission at high speed with the lowest cost due to utilizing the infrastructure that already exists. As a result of VLC's multiple features, it had been used in the indoor positioning system (IPS) in many pieces of research to obtain a high accuracy without conflict with the Radio Frequency (RF) spectrum besides it providing high security from any penetrating. However, achieving good performance parameters is essential and fundamental in evaluating the effectiveness of any indoor system that many pieces of research neglected and concentrated more on the positioning accuracy aspect. This paper investigates and analyses the performance of the indoor system that is designed and developed based on a VLC and proves its effectiveness through a comprehensive evaluation. Signal-to-noise ratio (SNR) and path loss are the performance parameters that are investigated in this system by varying the transmitted power, incidence angles, and Lambertian mode number in the line of sight (LOS) scenario. The examined system consists of one light-emitting diode (LED) in a 3D typical room and one photodiode (PD) in different ten locations along the half-diagonal line towards the corner. Utilizing a single LED in this system is important to avoid the interference that occurs when utilizing multiple LEDs; besides, it is more convenient in the indoor environment. The obtained results by MATLAB simulation show the reliable and effective performance of the proposed developed VLC system design by achieving a good SNR with low path loss. Furthermore, the proposed system approach proves how the optimum position of PD is crucial to obtain a strong signal with the lowest ratio of noise and losses.

## I. Introduction

The deployment of 5G opened the door to meet the incremental need for higher capacity in the wireless network imposed as a result of the rapid development of artificial intelligence (AI) and the internet of things (IoT) [1]. In spite of that, it is expected that 6G will exceed its need for the current wireless spectrum, which prompts the search for higher frequencies in terahertz to cover wider communication [2]. Visible light communication (VLC) considers a suitable and promising candidate technology that meets this expanding need of the wireless spectrum in the near future where its spectrum range of 400–800 THz [2]. It considers an integral part of optical wireless communication (OWC) that has high-speed connectivity and high data rates [3]. VLC has various features over other radio frequency (RF) technologies such as wireless fidelity (Wi-Fi), Bluetooth, and radio frequency identification (RFID) [4]. This green technology is useful in short-range scenarios where cannot penetrate the walls and this is extremely useful in terms of privacy issues [5]. As a health aspect, it has no harmful effects on health if compared to radio communications. This important feature highlights this technology to utilize in RF-restricted places like hospitals, laboratories, and airplanes due to it being interference-free to the RF [6,7]. From a cost aspect, it is efficient because its deployment depends on the infrastructure of the installed illumination with some additional devices.

Light emitting diode (LED) represents the source that is applied widely in VLC technology due to its numerous features that enable to get a reliable, solid, and efficient illumination system [8]. The main features of the LED can be represented as a long lifetime, low power consumption, lightweight, small size, high-lighting efficiency, low cost, safe to the human eye [9]. Also, it is easy to install, easy to use and can be modulated at higher data rates than conventional lighting sources [10]. The double function of the LED nominating it the best choice to utilize in homes and offices for communication as data transmission, the indoor positioning besides its basic function of illumination [9]. As an advanced step could be taken towards reducing power consumption radically, all the lights are replaced by the LEDs.

In this context, several research studies have appeared in the last decades that deal with indoor applications based on VLC [7][11,12]. The indoor positioning system (IPS) occupied a wide portion of the interest of researchers in these applications [13,14]. By applying different techniques based on VLC technology, a good level was achieved in the positioning of the receiver, reach to just a few centimeters from the actual position. For instance, 9.2 cm is the average positioning accuracy that is achieved based on the time-difference-of-arrival (TDOA) technique [15].

TABLE I COMPARISON BETWEEN DIFFERENT PROPOSED SYSTEMS THAT ARE BASED ON VLC-IPS AND USED DIFFERENT TECHNIQUES AND THEIR PARAMETERS PERFORMANCE

| Ref. | Year | Technique | System Model (m$^3$) | Experiment Type | Scenario | SNR (dB) | Path Loss (W) |
|---|---|---|---|---|---|---|---|
| [16] | 2017 | RSS | 5×5×3 | Simulation | LOS | 13.6 | — |
| [17] | 2018 | RSSI | 5×5×3 | Simulation | LOS | > 12 | — |
| [18] | 2018 | BPP | 5×5×3 | Simulation | LOS | 20.11 | — |
| [19] | 2020 | RSSI /TDOA | 5×5×3 | Simulation | LOS | 0.24×10$^{-3}$ | — |
| This work | 2023 | CSA-RSS | 5×5×3 | Simulation | LOS | 22.07 | 0.019 |

However, some of these researchers concentrate their investigation study to reach the precise positioning but they neglect the other important metrics such as in [15]. Some of them also took into their consideration calculating the received power besides the positioning. For instance, in [20] the author proposed using the Fingerprint technique based on VLC-IPS and the obtained received power was 0.92×10$^{-3}$ watts and 19.3 cm is a positioning error. In addition, some of them calculate the signal-to-noise ratio (SNR) such as in [16] where the proposed system based on VLC-IPS used the received signal strength (RSS) technique to get SNR value of 13.6 dB and 6 cm of the positioning error. The received signal strength indication (RSSI) technique was applied based on VLC-IPS in [17] where SNR obtained value was > 12 dB while the positioning error was <10 cm. While some took both, for instance in [19] hybrid technique RSSI / TDOA was proposed to apply in the system based on VLC-IPS to achieve better accuracy in the positioning error. It achieved 5.81 cm for positioning error and the obtained received power was 0.254×10$^{-3}$ watt besides the SNR obtained value was 0.24×10$^{-3}$ dB. In the research [18], the binomial point process (BPP) technique was proposed in the VLC system performance evaluation and the SNR performance value was 20.11 dB. Moreover, most of the research in this field neglected the path loss parameter. A comprehensive evaluation of any proposed system based on the important performance parameters considers significant and essential to prove its effectiveness and robustness.

Different proposed systems that are based on VLC-IPS and used different techniques had been compared with the new proposed system. The comparison was in terms of the used techniques, the system model that was applied, and the experiment type, scenario scheme, SNR, and path loss are given in Table I.

In this work, an indoor VLC system in the line of sight (LOS) model had been proposed and simulated to investigate and evaluate the performance parameters. Also, to find out how the light distribution impact against the photodiode's (PD) different locations on the system performance. This is an extension of our previous work [21], where a new technique for positioning was proposed named complementary and supplementary angles based on received signal strength (CSA-RSS). Excellent results were achieved in the average positioning error reached 4.2 cm and the received optical power was 4.5 watts. More precisely, the main contributions of this paper are outlined as follows:

- An indoor 3D system based on VLC technology is designed and developed in the LOS scenario because it covers most of the received power. A pair of LED-PD is used to investigate the effectiveness of the system through the performance parameters.
- Investigate the impact of PD movement from the room's floor center towards one of the corners with equal displacements in 10 locations against fixed LED in the ceiling's center on the obtained values of the SNR and the path loss parameters.
- Calculating the effect of varying the incidence angles, transmitted power, and Lambertian mode number against increasing the distance between the LED-PD.
- Evaluating the results for different configurations by MATLAB R2019a simulation of the proposed system.

The content of this paper is organized as follows; section II illustrates the model of the proposed system. Section III investigates and demonstrates the performance of the important parameters of the proposed system based on the simulation results with the evaluation and discussion. Finally, this work had been summarized in Section IV.

## II. SYSTEM MODEL

The proposed model of the indoor system based on VLC is shown in Fig. 1, a standard-size empty room with dimensions of 5 m × 5 m × 3 m. LOS is the proposed model of OWC to investigate the performance of the system based on the fundamental parameters for evaluation. The system consists of a single LED as an access point located in the center of the room's ceiling to achieve a uniform distribution of light and obtain an optimum LOS that covers most points of the room for the proposed indoor environment. The coordinate of the LED is (2.5, 2.5, 3) m where the distance is equal between the LED and all wall edges. A single PD that changes its place from the room's floor center towards one of the corners in a half diagonal line with equal displacements in 10 locations as presented in Fig. 1. The coordinates of the PD's ten locations are (2.50, 2.50, 0), (2.23, 2.23, 0), (1.96, 1.96, 0), (1.69, 1.69, 0), (1.42, 1.42, 0), (1.15, 1.15, 0), (0.88, 0.88, 0), (0.61, 0.61, 0), (0.34, 0.34, 0), (0.07, 0.07, 0) m. An investigation had been made to explore the effect of increased linked distance between LED-PD dramatically on the system performance. The LED broadcasts the visible light signals which are transmitted with a unique location code based on its location coordinates as well as its lighting function.

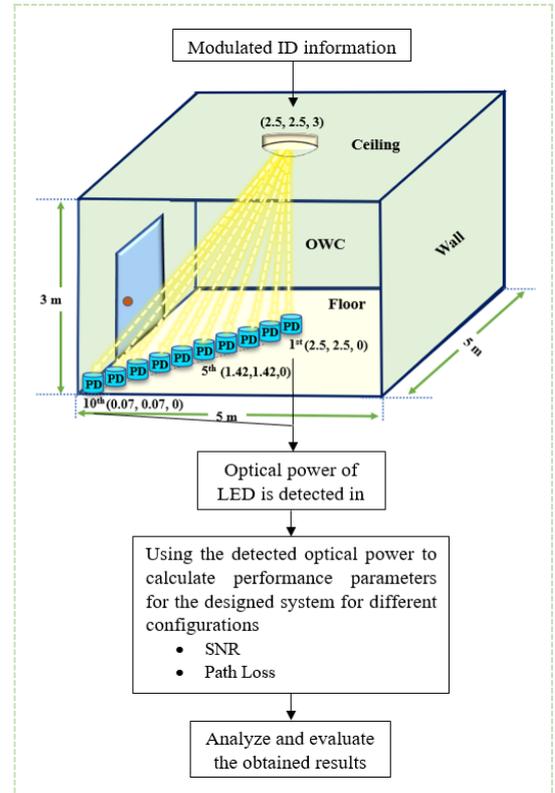

Fig. 1. Scheme of the indoor system based on VLC

A. *The proposed scenario's configuration*

- In the LOS scenario of the indoor proposed system model that is based on VLC. One LED as an access point for the entire room is fixed in the center of the ceiling with the coordinates (2.5, 2.5, 3) m to reach optimum distribution which almost covers most room points symmetrically. The dimensions of the typical empty room are 5 m × 5 m × 3 m. The impact of increasing the distance due to the PD movement in 10 locations on the floor plane against the fixed LED reflects on the whole system's performance.

- At the first location with coordinate (2.5, 2.5, 0) m, the PD was placed directly under the LED intensity with a straight line of LOS where 3 m is the distance between LED-PD.

- The proposed movement of the PD is taken place symmetrically in the *X* and *Y* planes with equal displacements between the positions starting from the center toward the one of corners.

- The PD in its 10 locations moves in a half-diagonal line on the floor plane until the tenth position at the corner with the coordinate (0.07, 0.07, 0) m.

B. *VLC channel model and its Geometry*

The indoor VLC proposed system consists of a LED as a transmitter that transmits the visible light signals to a 3D empty room of size 5 m × 5 m × 3 m and a PD as a receiver that is placed on the floor plane for signals receive. The LED is mounted in the middle of the room roof with coordinates (2.5, 2.5, 3) m at a height of 3 m and the distance between it and the four edges of the walls is 2.5 m. The LOS link model is the optical channel considered in this investigation study because it comprises most of the received power. The signals are transmitted in a direct path through the LOS channel to reach the PD that receives and measures the strength of this optical power. The ten locations of the PD to investigate their effect are (2.50, 2.50, 0), (2.23, 2.23, 0), (1.96, 1.96, 0), (1.69, 1.69, 0), (1.42, 1.42, 0), (1.15, 1.15, 0), (0.88, 0.88, 0), (0.61, 0.61, 0), (0.34, 0.34, 0), (0.07, 0.07, 0) m. These ten locations start from the room center in a direct LOS with LED intensity until the tenth location at the corner. The geometry of the VLC channel model between the LED and the ten locations of the PD in the LOS link is shown in Fig. 2 where $d_{LED-PD}$ is the distance between LED-PD and $d_{vertical}$ is the vertical distance between the ceiling and floor planes. Also, the irradiance angle $\phi_{irr}$, incidence angle $\theta$, the LED's half-power angle ($\phi_{1/2}$), and the field of view (*FOV*) angle of the PD are shown in Fig. 2 where these angles play important role in the findings obtained.

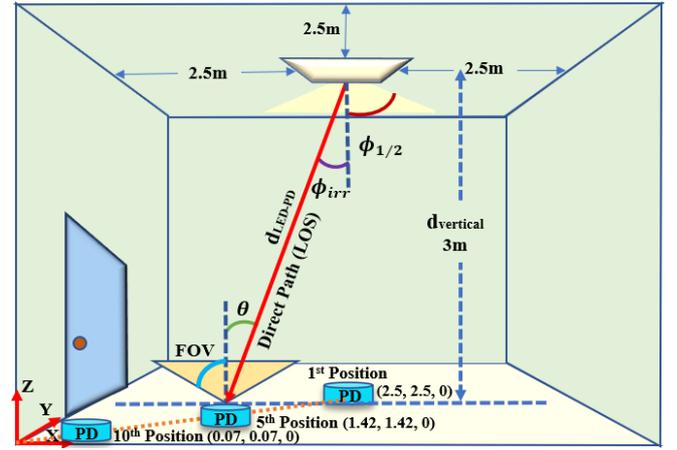

Fig. 2. VLC channel model geometry

*1) Signal to noise ratio*

In order to obtain outstanding performance of the VLC system and achieve higher capacity and findings, SNR plays a major and important role. This is achieved by SNR improvement which is reflected in the rapid future development of indoor systems. One of the parameters that have a negative effect on the received power ($P_{Received}$) is noise where $P_{Received}$ is affected by shot and thermal noises. Shot noise is the fluctuating that occurs because of the incident optical powers for the desirable and the ambient light sources. Thermal noise is the PD's fluctuating that is caused by temperature changes of the electric circuit in the receiver. The overall noise variance is *N* and usually is modelled as the additive white Gaussian noise (AWGN) and it is given as [22]:

$$N = \sigma_{Thermal}^2 + \sigma_{Shot}^2 \quad (1)$$

SNR plays a significant role to attain effective communication and examine any proposed system. SNR is more related to $\sigma^2_{Shot}$ where it can be expressed as [23]:

$$SNR = 10\log_{10} \frac{(P_{Received} R_P)^2}{\sigma_{Shot}^2} \quad (2)$$

Where $R_P$ denotes the PD's responsivity, the variance of the shot noise can be expressed as [24]:

$$\sigma_{Shot}^2 = 2qR_P P_{Received} B + 2qI_{bg} I_2 B \quad (3)$$

Where *q* is the electron charge, *B* is the equivalent noise bandwidth, $I_{bg}$ is the background current, and $I_2$ is a noise bandwidth factor.

The SNR value can give a precise reflection on the quality of the received signal, as it is affected by the value of the $P_{Received}$ as an important parameter in its equation where it had been calculated based on the RSS-CSA technique that is proposed and proved based on the obtained good results previously in our work [21] and can be expressed as follow [4]:

$$P_{Received} = \frac{P_{trans}}{(d_{LED-PD})^2} f(\phi_{irr}) A_{eff}(\theta) \quad (4)$$

where $P_{trans}$ represents the transmitted power from the LED in the VLC system, $f(\phi_{irr})$ is the intensity of the radiant angle of the LED's irradiant angle ($\phi_{irr}$), $d_{LED-PD}$ is the distance between the LED-PD, and $\theta$ represents the incidence angle which changes based on the PD's locations as shown in Fig. 2. $A_{eff}$ is the effective area in the PD that combines the transmitted signals and can be expressed as follows [4][12]:

$$A_{eff}(\theta) = \begin{cases} Ah(\theta)g(\theta)\cos\theta, & \theta \leq FOV \\ 0, & \theta > FOV \end{cases} \quad (5)$$

*A* symbolizes the PD's surface area, the field of view of the PD can be denoted as *FOV*, $h(\theta)$ expresses the gain of the optical filter while the gain of the concentrator is represented as $g(\theta)$ [25]:

$$g(\theta) = \begin{cases} \dfrac{n^2}{\sin^2(FOV)}, & 0 \leq \theta \leq FOV \\ 0, & \theta > FOV \end{cases} \quad (6)$$

The concentrator's refractive index expressed as *n* [24][26].

Typically, the power distribution of the LED's profile is modeled as a Lambertian emission, and $f(\phi_{irr})$ can be denoted as follows: [4]:

$$f(\phi_{irr}) = [\dfrac{(m+1)}{2\pi}]\cos^m\phi_{irr} \quad (7)$$

The value of *m* refers to the directivity of the LED and represents the LED's Lambertian model number. Also, it is linked to the LED's half-power angle $\phi_{1/2}$ and can be expressed as [4][27]:

$$m = \dfrac{-\ln(2)}{\ln(\cos(\phi_{1/2}))} \quad (8)$$

The distance between LED-PD in the LOS channel model can be calculated as follows [28][29]:

$$d_{LED-PD} = \sqrt{(X_i - X_j)^2 + (Y_i - Y_j)^2 + (Z_i - Z_j)^2} \quad (9)$$

Where $(X_i, Y_i, Z_i)$ represent the LED location which is fixed and known as (2.5, 2.5, 3) m While $(X_j, Y_j, Z_j)$ represent the PD's ten locations. Table II illustrates the calculated distance between the LED and the PD's ten locations based on equation 9. Its movement occurs symmetrically in the *X* and *Y* axes starting from the center of the floor plane toward one of the corners.

TABLE II. COORDINATES AND DISTANCE BETWEEN LED AND PD'S TEN LOCATIONS

| LED's Coordinate | PD's Coordinates | | Distance |
|---|---|---|---|
| 2.5, 2.5, 3 | 1st/At center | 2.5, 2.5, 0 | 3 m |
| | 2nd | 2.23, 2.23, 0 | 3.024 m |
| | 3rd | 1.96, 1.96, 0 | 3.095 m |
| | 4th | 1.69, 1.69, 0 | 3.211 m |
| | 5th | 1.42, 1.42, 0 | 3.366 m |
| | 6th | 1.15, 1.15, 0 | 3.555 m |
| | 7th | 0.88, 0.88, 0 | 3.774 m |
| | 8th | 0.61, 0.61, 0 | 4.017 m |
| | 9th | 0.34, 0.34, 0 | 4.281 m |
| | 10th/At corner | 0.07, 0.07, 0 | 4.561 m |

*2) Path loss*

Path loss is one of the significant metrics of the wireless channel evaluations for any proposed system. It is used to compute the decrease in the power of a signal as it propagates away from the transmitter (LED). Path loss is a major component in the analysis and design of the link in a telecommunication system where it plays a vital role in wireless network planning. Free space path loss is the proposed module in this indoor VLC system where it computes the loss of signal power in a LOS propagation path without any reflections or shadowing, and can be expressed as [30][31]:

$$P_{Loss} \approx \dfrac{(m+1)A}{2\pi(d_{LED-PD})^2}\cos^m(\phi_{irr})\cos(\theta) \quad (10)$$

The system simulation parameters values for different configurations which took into consideration in this research paper are given in Table III.

TABLE III. SIMULATION PARAMETERS OF PROPOSED INDOOR VLC SYSTEM

| Parameter | Value |
|---|---|
| Room Dimensions | 5 m × 5 m × 3 m (Length × Width × Height) |
| PD's surface area | 2.25 mm² |
| LED's transmitted Power | (8, 10, 12, 15) Watts |
| Gain of the optical filter $h(\theta)$ | 1.0 |
| Reflective index (*n*) | 1.5 |
| Lambertian model number (*m*) | 1.3 |
| LED's half-power angle | 60° |
| Field of view (*FOV*) | 90° |
| Incidence angle $\theta$ | 60°, 70°, 80°, 90° |
| PD responsivity ($R_P$) | 0.6 A/W |
| **Noise Parameters** | |
| Background current ($I_{bg}$) | 5.1 mA |
| Equivalent noise bandwidth (*B*) | 50 MHz |
| Noise bandwidth factor ($I_2$) | 0.562 |
| Electron charge (*q*) | 1.6×10⁻¹⁹ coulombs |

III. SIMULATION RESULTS AND DISCUSSION

In this section, the simulation results by MATLAB R2019a that investigate the proposed indoor VLC system performance are discussed and analysed in detail in the LOS scenario. Different configurations of PD locations and how significantly affect the parameters' performance are demonstrated. All the required parameter values for the simulation are given in Table III. The aim is to investigate and prove the system's performance effectiveness in a comprehensive manner that reflects on the indoor systems' future. Also, it keeps pace with rapid development by achieving high speed in the transmission of data with the lowest cost. Fig. 1 and Fig. 2 illustrate the distribution scenario of a single LED and PD's ten different locations in a 5 m × 5 m × 3 m room environment. To achieve a uniform distribution of LED lighting and good coverage for most of the points, the LED is mounted in the middle of the room's ceiling. The different locations of the PD through its movement reflect a clear description of how the system performance is affected dramatically by this movement away from the LED intensity. Equal displacements between every location of the PD's ten locations on the *X* and *Y* axes in a symmetrical way enable regular evaluation of system performance.

### A. Signal to Noise Ratio performance

In order to obtain a premium performance of the indoor VLC system, SNR has a key role that influences achieving that. SNR profile in the room is changed based on the locations of the LED-PD of the different arrangements. Therefore, SNR has to be calculated to guide the optimum LED light distribution inside the indoor environment. To investigate the SNR performance with the design parameters based on equation 2, several values of the incidence angles and the transmitted power had been selected to examine its performance and robustness as analysed and discussed in the figures below.

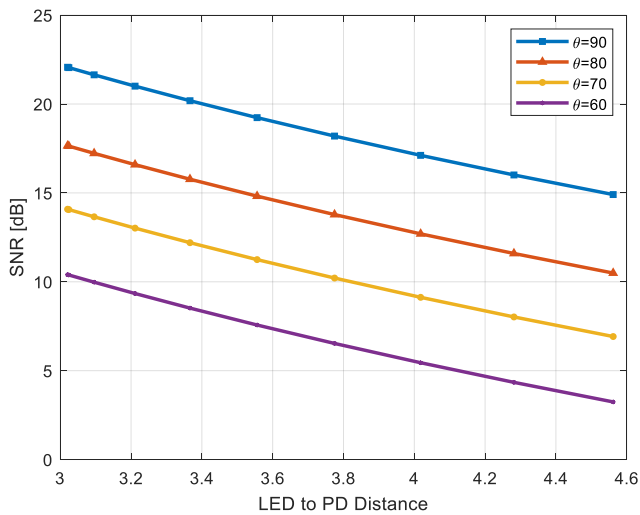

Fig. 3. Performance of SNR versus the distance between the LED-PD with different incidence angles.

Fig. 3 shows the performance of the SNR versus the distance between LED-PD with several proposed incidence angles of 60°, 70°, 80°, and 90°. Obviously, the obtained SNR value decreases gradually as long as distance increases in the PD's movement through all the selected incidence angles. In this context, the maximum SNR value is achieved at the PD's first location where the distance is the shortest path between LED-PD of 3 m while the minimum value is obtained at the longest distance of 4.56 m at the corner. Based on equation 2, the SNR value is impacted directly by the received power value and is considered a major parameter that affects the obtained results influentially. Also, the incidence angles are considered one of the effective and influential parameters in the design of a successful indoor VLC system. Therefore, there is a noticeable difference in SNR values upon each of the selected incidence angles. For instance, 22.07 dB is the maximum value of SNR achieved at 90° due to the strongest LOS between the center of LED intensity and the first PD's location at the room's floor center at a distance of 3 m. On the other hand, the minimum value of SNR obtained at 60° was within 10.40–3.24 dB. In the case of the incidence angles of 70° and 80°, it is obvious the performance of the SNR value shows a decrement of 14.08–6.92 dB and 17.63–10.49 dB respectively. The difference between the maximum SNR obtained values at 90° and 60° is 11.67 dB which refers to a major and clear difference and confirms the significance of the incidence angles' role in the results. In conclusion, based on the proposed configurations and the obtained results, it is observable that the SNR value reduces with the incidence angle value decrease versus the incremental in LED-PD distances directly due to being away from the LOS channel of the LED source toward one of the corners.

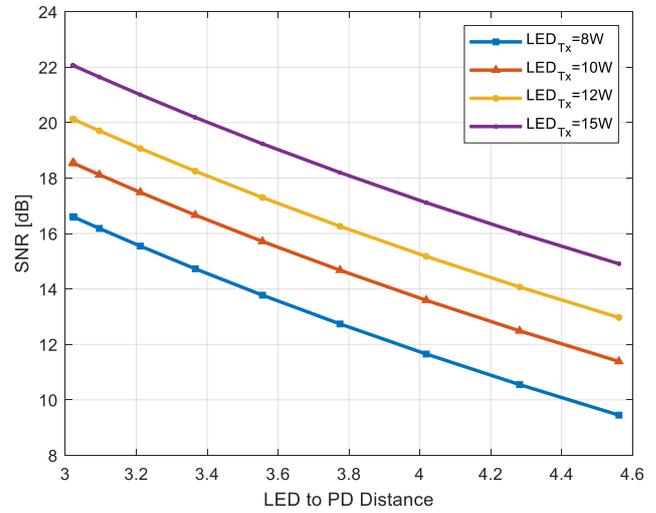

Fig. 4. Performance of the SNR versus distance between the LED-PD with different values of the transmitted power.

Fig. 4 shows the SNR performance versus the distance between LED-PD in different configurations with various proposed values of the transmitted power of 8, 10, 12, and 15 watts. The performance of SNR witnesses a noticeable decrease with the increase of the distances for the varied transmitted power values. The transmitted power is considered one of the significant design parameters where it effects the received power according to equation 4 and any change in the received power reflects on the obtained SNR value based on equation 2. In this context, it is noticeable that the maximum SNR values are achieved at the PD's first position where the distance between LED-PD is 3 m. The minimum SNR values are obtained at the PD's tenth position at one of the corners where the distance between LED-PD is 4.56 m. Based on the obtained results, the maximum achieved value of the SNR is 22.07 dB in the PD's first location at the transmitted power of 15 watts and this is attributed to the PD being closer to receiving the strongest signal from the LED source. In this regard, the minimum value of the SNR was decreased from 16.61 to 9.45 dB at the transmitted power of 8 watts. For the proposed transmitted power of 10 and 12 watts, the SNR value is increased to record 18.55 dB and 20.13 dB respectively at the first location of the PD. On the other hand, the distance between LED-PD increases due to PD's moving far away from the LED source intensity toward the corner, SNR value decrease to record 11.39 dB and 12.97 dB respectively. By comparing the maximum obtained values of SNR at the transmitted powers of 15 and 8 watts, the difference is 5.46 dB which refers to the effect of the transmitted power as an important design parameter on the findings. Comparing the difference of the maximum obtained values of SNR of the incidence angle with 11.67 dB, and the transmitted power of 5.46 dB, it is obvious that the incidence angle has the biggest impact on the results. In conclusion, based on the obtained results, it is observable that the SNR value increases with the incremental in the transmitted power values but it decreases with the increase in LED-PD distances because of moving far from the direct LOS channel with the LED intensity toward the corner.

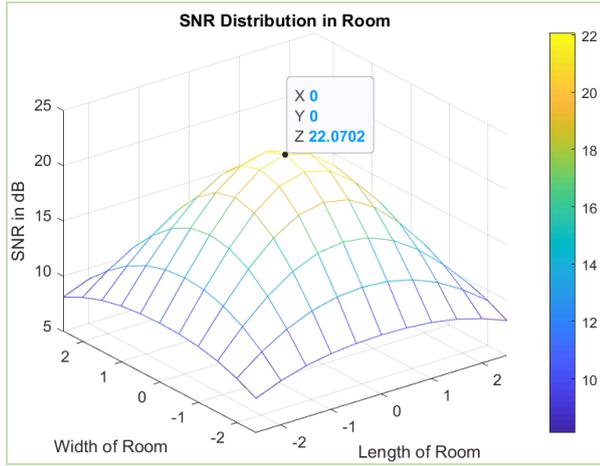

Fig. 5. 3D distribution of SNR inside the proposed room.

To take a deeper look, Fig. 5 illustrates the distribution of SNR inside the proposed room in a 3D model. As shown, 22.07 dB is the maximum value obtained from SNR where the PD in the first location lies directly under the LED source radiation. For the design parameters applied values, the incidence angle is 90° and the value of the transmitted power is 15 watts.

*B. Path loss performance*

In many research studies were conducted on indoor VLC systems, the path loss parameter had not been taken into consideration widely as the other parameters. Whereas it is important for any proposed system to give a comprehensive evaluation to achieve an effective design by proving its robustness for the indoor environment. To investigate the path loss performance with the design parameters based on equation 10, several values of the incidence angles had been selected to examine their effect on the proposed system. Also, the Lambertian mode number ($m$) has a significant influence on the obtained results as it refers to the directivity of the LED and is linked to the LED's half-power angle. The values of the shot noise and other design parameters of this research paper are given in Table III and the distances are calculated according to equation 9 and given in Table II. The path loss performance was analyzed and discussed in the figures below.

In Fig. 6, the path loss performance versus the distance between the LED and different locations of the PD with different incidence angles 60°, 70°, 80°, and 90° are investigated. Based on the obtained results and according to equation 10, the path loss value increases simultaneously as long as distance increases in the PD movement far from the LED. In this context, the lowest path loss value is achieved when the PD lies in the first location in the direct LOS with the LED radiation at the shortest distance of 3 m. On the other hand, the highest value of the path loss is obtained at the corner at the longest distance of 4.56 m. In consideration of the proposed incidence angles, it can be observed that the lowest value of the path loss is obtained at 90° while the highest is at 60°. This is because at 90◦ the PD lies directly under the LED source radiation which leads to a low path loss value, but this value increases when the incidence angle is reduced due to being far from the LED and the LOS channel. For instance, 0.008 watts is the lowest obtained value of the path loss at 90° at 3 m due to being in a strong LOS with direct LED intensity. With PD moving towards the corner, the path loss continues to increase until reaching the tenth location to record 0.019 watts as the highest path loss value at 4.56m. This is because the LOS becomes weak as a result of being away from the LED radiation. In this context, the highest value of the path loss obtained at 60◦ was within 0.041–0.093 watts because of the weak occurs in the signal versus the increased noise far from the direct LOS with the LED intensity. In the proposed incidence angles case of 70◦ and 80◦, it is noticeable that the performance of the path loss value shows an increment of 0.029–0.067 watts and 0.018–0.041 watts respectively. Calculating the difference between the highest path loss obtained values at 60◦ and 90◦ is 0.074 watts which indicates a clear difference. This reflects on the obtained signal and emphasizes the important role of the incidence angles in the findings. In conclusion and based on the proposed scenario and obtained results, the path loss value increases with the distance between LED-PD increases, and versus decrease in the incidence angles.

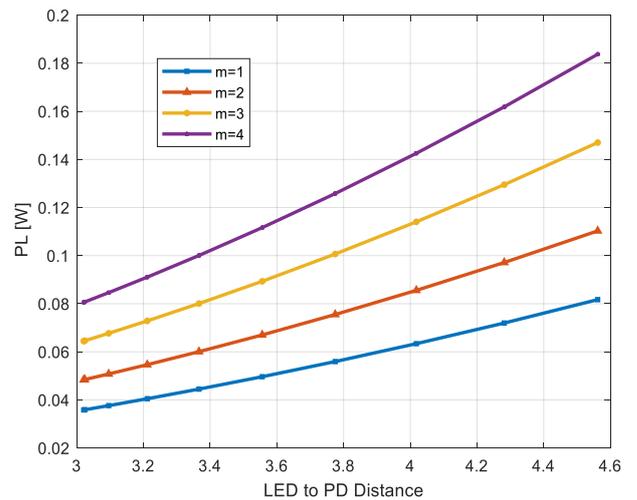

Fig. 7. Performance of Path Loss versus the distance between the LED-PD with different Lambertian mode number.

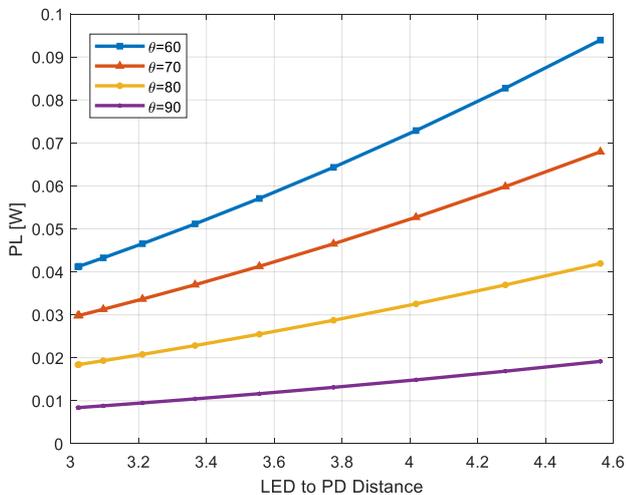

Fig. 6. Performance of Path Loss versus the distance between the LED-PD with different incidence angles.

in Fig. 7, the path loss performance was examined versus the distance between LED-PD with different Lambertian Mode Numbers $m$=1, 2, 3, and 4. According to equation 10 and upon the achieved results, there is an increment in the path loss with distance increases because of PD movement away from the LED and this causes a loss in the signal strength. In this context, $m$ plays a major role in the increased values of the path loss significantly due to the light will be more concentrated in the center where it refers to the LED's half-power angle. Obviously, the highest path loss value has been obtained at the tenth location of the PD at the corner with a distance of 4.56 m and $m=4$. On the other hand, the lowest value of the path loss was recorded at the first location of the PD at the floor center which lies in the direct LOS channel with a distance of 3 m and $m=1$. With PD movement far from the LED, the path loss value increases to record the highest value at $m=4$ in the range of 0.080–0.183 watts. At $m=1$ the minimum proposed values, the light concentration at the center becomes less where the lowest recorded value of the path loss was within 0.035–0.081 watts. In the case of $m=2$ and $m=3$, it is notable that the performance of the path loss value witnesses an increase of 0.048–0.110 watts and 0.064–0.147 watts respectively. Finding the difference between the highest obtained values at $m=4$ and $m=1$ is 0.102 watts. This difference value indicates the important role of $m$ that affect the results significantly. In conclusion, the path loss increases when the distance between LED-PD increases simultaneously. By increasing the $m$ value, the light becomes more concentrated and this reflects to clear increase in path loss value. Comparing the obtained different results based on varying the incidence angle, and $m$ as design parameters, it is obvious that $m$ has the biggest effect on the path loss increment.

IV. CONCLUSION

After the 5G deployment, meeting the growing demand to achieve high capacity in the communication field appears clearly to be the answer for the near future to keep pace with the occurring rapid development to cover wider communication. The several distinguished features of the VLC technology make it the optimum candidate for safe and secure indoor communication to meet this increased need.

However, due to a lot of the VLC's advantages over other technologies, many studies and research had been conducted to investigate the VLC system performance in the indoor environment in the positioning and other services but some of them were not interested in fully investigating the proposed systems. This is achieved by examining several important performance parameters along with positioning accuracy to obtain a comprehensive evaluation of any proposed indoor system.

This paper proposes to design and develop an effective 3D indoor system based on VLC; it consists of a pair of LED-PD inside a standard room of 5 m × 5 m × 3 m. This work investigates and analyzes the proposed system performance in terms of SNR and path loss. A comprehensive evaluation had been made by examining the effect of the crucial design parameters of the transmitted power, incidence angles, and Lambertian mode number on the performance parameters and is an expansion of our previous investigation work on indoor positioning and the received power. This approach deals with investigating the PD optimum location to achieve the strongest signal and fewer losses. The achieved findings by MATLAB R2019a simulation prove the effective and credible performance of the proposed indoor system.